\documentstyle[aps,prb,preprint]{revtex}
\begin{document}
\draft
\title{Strongly Anisotropic Transport in Higher Two-Dimensional Landau Levels}
\author{R. R. Du$^{a,b}$, D. C. Tsui$^{a,c}$, H. L. Stormer$^{d,e}$,
L. N.  Pfeiffer$^d$,\\ K. W. Baldwin$^d$, and K.W. West$^d$}
\address{a) Dept. Electr. Eng., Princeton University, Princeton, NJ
\\b) Dept. Phys., University of Utah, Salt Lake City, UT \\c)
Dept. Phys., Princeton University, Princeton, NJ \\d) Bell Labs,
Lucent Technologies, Murray Hill, NJ \\e) Dept. Phys. and
Dept. Appl. Phys., Columbia University, New York, NY}
\date{November, 9 1998}
\maketitle

\begin{abstract}
Low-temperature, electronic transport in Landau levels $N>1$ of a
two-dimensional electron system is strongly anisotropic. At half-filling of
either spin level of each such Landau level the magnetoresistance either
collapses to form a deep minimum or is peaked in a sharp maximum, depending
on the in-plane current direction. Such anisotropies are absent in the $N=0$
and $N=1$ Landau level, which are dominated by the states of the fractional
quantum Hall effect. The transport anisotropies may be indicative of a new
many particle state, which forms exclusively in higher Landau levels.
\end{abstract}


Two-dimensional electrons in high magnetic field exhibit a multitude of
novel electronic phases\cite{ref1,ref2}. The electronic states of the
fractional quantum Hall effect (FQHE) are the most abundant and probably
also the best known. Most of the transport studies have been performed on
the lowest Landau level, $N=0$. Magnetotransport on higher Landau levels has
been limited to the still enigmatic even-denominator state at filling factor 
$\nu =5/2$ and features in its immediate vicinity in the $N=1$ Landau level%
\cite{ref3,ref4,ref5,ref6}. Still higher Landau levels $(N>1)$ are largely
uncharted from the point of view of novel electronic phases.

With our high-mobility $(\mu =1.2\times 10^7cm^2/V\cdot sec)$ sample of
density $n=2.3\times 10^{11}cm^{-2}$ we were able to pursue transport
features at partial fillings of higher Landau levels at magnetic fields as
low as $1.5T$. Beyond the previously observed minima in the
magnetoresistance $R_{xx}$ at $\nu =5/2$ and $7/2$\cite{ref3,ref4,ref5,ref6}
we find strong minima at half-filling in higher Landau levels, such as $\nu
=9/2,11/2,13/2,15/2$ and $17/2$. These new features are flanked by other
minima, whose positions do not seem to follow the sequence observed in the
lower Landau levels. Most remarkably, the magnetoresistances at exactly
half-filling from $\nu =9/2$ through $\nu =17/2$ are strongly dependent on
the current-voltage contact configuration. Minima at these positions turn
into steep maxima after rotating the current direction by 90$^{\circ }$.
This puzzling observation has been reported earlier in a preliminary
communication\cite{ref7} and has recently also been observed by Lilly {\em%
et al}\thinspace\cite{ref8}. The origin of these transport features in higher Landau
levels and the reason for their extreme anisotropy remain unresolved but may
be the result of the formation of electronic phases different from the
well-known FQHE states.
 
Our square specimen of $5mm\times 5mm$ was cleaved
along the $[\overline{1}10]$ and $[1\overline{1}0]$ directions from an MBE
grown wafer. Eight indium contacts were diffused symmetrically around the
edges of the sample. Experiments were performed in a dilution refrigerator
immersed into a superconducting solenoid. Magnetoresistance, $R_{xx}$, and
Hall resistance, $R_{xy}$, traces were recorded using a standard
low-frequency lock-in technique with a $1-2nA$ excitation current. The
magnetic field direction was always kept perpendicular to the 2D-electron
system.

Fig.1 shows two magnetoresistance traces recorded at $T=60mK$ between Landau
level filling factor $\nu =2$ and $\nu =12$, {\em i.e.} for $N=1$ through $6$. 
In the $N=1$ Landau level, between $\nu =2$ and $4$, the previously observed
characteristic features of the FQHE are well developed. In particular, the
minima of the even-denominator FQHE states at $\nu =5/2$ and $\nu =7/2$ are
clearly visible. Quite remarkably, however, in this high-mobility specimen
features in $R_{xx}$ , not unlike FQHE features, are discernible up to $\nu
\sim 9$, which corresponds to the $N=4$ Landau level. In particular, as seen
in the lower trace $(B)$, minima at half-filling are unexpectedly strong and
visible up to $\nu =17/2$. In the $N=2$ Landau level their strengths even
exceeds those of the $\nu =5/2$ and $7/2$ minima. This is a surprising
result, since the latter reside at higher magnetic field, where FQHE
features are generally better developed. In fact, while there is a clear,
progressive increase in strength from $\nu =17/2$ to $9/2$, the $\nu =7/2$
and $5/2$ minima, much weaker than the $9/2$ minimum, seem to interrupt this
trend.

Beyond the strong minima at half-filling, additional satellite minima are
visible in the $N=2$ Landau level. It is difficult to associate these
structures with a particular rational filling factor, but they are located
in the vicinity of $1/4$ and $3/4$ filling. Similar to the situation at
half-filling, these satellites appear much more dominant in the $N=2$ Landau
level than the satellites around $\nu =7/2$ and $5/2$, which are most likely
associated with $1/3$ and $2/3$ fillings of the $N=1$ Landau level. Already
at this stage one wonders whether the new features for $N>1$ are at all
related to the minima in the $N=1$ and $N=0$ Landau level, which are of FQHE
origin.

The most striking discrepancy between the $N>1$ and $N\leq 1$ data lies in
their extraordinary dependence on the in-plane current direction. The top
trace $(A)$ of Fig.1 shows data from the same specimen at the same
temperature as the bottom trace $(B)$. The only difference is the
configuration of the current and voltage probes, which have been rotated by $%
90^{\circ }$ from $B$ (see inserts to Fig.1). While the $N=1$ FQHE
structures are essentially independent of the in-plane current direction,
the {\em minima} at half-filling in higher Landau levels turn into a set of
sharp {\em peaks}. In particular the spike at $\nu =9/2$ exceeds all other
features of the spectrum by almost a factor of two. Towards higher filling
factors (smaller $B$-fields) the strength of the peaks decreases. An
alternating strong-weak-strong pattern seems to be superimposed, possibly
being associated with the alternating spin polarization of the electron
system. Further inspection reveals that the data in trace $A$ of Fig.1 seem
to be a superposition of trace $B$ and these sharp spikes at half-filling
for $N>1$. This becomes particularly apparent around $\nu =9/2$ and $11/2$,
where the satellite minima of the lower trace are preserved and new, inner
satellites have been created by the addition of the peaks to the previous
minima at half-filling. This notion is supported by data taken at $25mK$ in
the configuration $A$ where only the peaks remain and all other features
have totally disappeared for $N>1$ (see insert trace $C$).

The data of Fig.1 are very puzzling. Such extreme anisotropies are absent at
filling factors $\nu <4$, {\em i.e.} $N<2$, in any two-dimensional specimen we
have ever studied. On the other hand, we have now observed the anisotropy
seen in Fig.1 for $\nu \geq 4$ in several other samples of electron
densities $1.1$ to $2.9\times 10^{11}$ $cm^{-2}$ and mobilities $6.8$ to $%
12.8\times 10^6cm^2/V\cdot sec$. From this we conclude, that anisotropic
transport is a pervasive feature around half-filling in Landau levels $N>1$.

To further investigate the phenomenon we measured the temperature dependence
of $R_{xx}$ around $\nu =9/2$ and $11/2$, where in Fig.1 the most pronounced
anisotropies occur, and contrast it with the pattern around $\nu =7/2$. As
seen in the bottom trace of Fig.2, $R_{xx}$ is essentially isotropic at $%
115mK$ from $\nu =3$ to $\nu =6$. This isotropy largely persists over most
of the filling factor range down to the lowest temperatures. However, at $%
\nu =9/2$ and $\nu =11/2$ the traces bifurcate into a rapid rise and rapid
drop, depending on the in-plane current direction. Fig.3 quantifies this
behavior, showing the $T$-dependence at $\nu =9/2$ for both current
directions on an Arrhenius plot. After similar high-temperature values the
peak shows a very rapid, approximately exponential rise over more than one
order of magnitude while the minimum experiences an equivalent exponential
drop. The characteristic energies, $E$, for the exponential part of the $T$%
-dependencies, are very similar $E\sim 0.55K$, where $R_{xx}\propto exp(\pm
E/kT)$. Both traces saturate at temperatures $T<\sim60mK$. In fact, the $T$%
-dependencies for the maximum and minimum are almost mirror symmetric,
suggesting the same phenomenon to be responsible for the appearance of
either transport behavior. The saturation value of the maximum of $%
R_{xx}\sim 1.5k\Omega $ is close to the resistance step between the
neighboring $\nu =5$ to $\nu =4$ Hall plateaus of $\Delta
R_{xy}=(1/4-1/5)h/e^2\sim 1.3k\Omega $. Inspection of our lowest temperature
data at $25mK$, shown in the inset to Fig.1, reveals that such a relation
between low-temperature $R_{xx}$ peak value and step between neighboring $%
R_{xy}$ plateaus holds, to within $25\%$, for all the spikes observed in $%
R_{xx}$. 

The large anisotropies in $R_{xx}$ are limited to $N>1$ and absent
in the $N=1$ Landau level. This is clearly seen in the field range around $%
\nu =7/2$ in Fig.2. Some discrepancies, particularly in the height of the
flanks close to $\nu =3$ and $4$, are discernible and are found in most
other 2D transport experiments. While their specific origin is also unclear,
they never have the strong attribute of turning a minimum into a maximum or
vice versa as it is observed around half-filling for $N>1$. In fact, the
strongly anisotropic behavior seems to be limited to the immediate vicinity
of half-filling. This can be seen in Fig.2, where the flanks of each band
around $\nu =9/2$ and $11/2$ show little angular dependence. In particular,
the minima flanking the features at half-filling, indicated by vertical,
dashed lines, show isotropic transport behavior. As an example for this lack
of anisotropy the right panel of Fig.3 quantifies the temperature dependence
of the minimum at $\sim 2.22T$ in Fig.2 in a standard Arrhenius plot for
both contact configurations. Both show the same exponential {\em drop} in 
$R_{xx}$ with an activation energy of $\Delta =1.7\pm 0.2K$, where 
$R_{xx}\propto exp(-\Delta /2kT)$. This obviously isotropic behavior is in 
stark contrast to the strongly anisotropic $T$-dependence at $\nu =9/2$ in the
left panel of the same figure. 

Compared to $R_{xx}$, the Hall resistance, $R_{xy}$, is
only slightly anisotropic as seen in the top traces of Fig.2. At the highest
temperature of $115mK$ (dashed trace) the Hall resistance from both current
directions practically coincide (only one is shown). At the lowest
temperature of $60mK$ the Hall resistance remains isotropic over most of the
field range (solid and dotted line) with a noticeable stretch of anisotropy
around $\nu =9/2$ and $11/2$. We have not further investigated this
anisotropy in the Hall resistance, since it could well arise from a small
admixture of the strongly anisotropic $R_{xx}$ into $R_{xy}$. While most of
the Hall resistance is well behaved, an unusual observation is made in $%
R_{xy}$ at the position of the satellite minima, indicated by the vertical,
dashed lines in Fig.2. They are located at $\nu =5.76,5.26,4.73$ and $4.28$,
close to quarter-filling by either electrons or holes, although it is
unclear whether they can be labeled with such rational fractions. Features
around such even-denominator filling factors, observed earlier in the lowest
Landau level, eventually developed into nearby odd-denominator FQHE states%
\cite{ref9,ref10}. The $R_{xy}$ values of such earlier states initially were
intermittent, but eventually approached the correct, quantized value of the
nearby odd-denominator state. The satellite minima highlighted in Fig.2 show
quite a different $R_{xy}$ development. While deep minima are developing in $%
R_{xx}$ at these filling factors, the concomitant $R_{xy}$ values do not
tend towards the usual fractional Hall plateaus which are cutting through
the classical Hall line. Instead, in spite of FQHE-like minima in $R_{xx}$ ,
their $R_{xy}$ values converge towards the nearest {\em integer} quantum
Hall plateau. This behavior is particularly well developed at $\nu \sim 4+
1/4$ $(B\sim 2.22T)$ but is also apparent at the other ``quarter-filling''
factors for $N>1$ (vertical, dashed lines). 

Our central experimental finding
can be summarized as follows: Electronic transport in the higher Landau
levels, $N>1$, differs in several ways from the usual electronic transport
in Landau levels $N\leq 1$. Whereas in the lowest Landau levels the standard
features of the FQHE dominate, transport around half-filling in higher
Landau levels is extremely anisotropic, developing large maxima or deep
minima depending on the in-plane current direction. Satellite minima in the
vicinity of quarter-filling do not show this anisotropy, but forgo FQHE
plateau formation in $R_{xy}$, approaching instead IQHE values for their
concomitant plateaus.

The observed anisotropy can be of intrinsic or extrinsic origin. Two
possible extrinsic sources come to mind: misalignment of the interface from
the desired $[001]$ direction or an in-plane electron density gradient. We
have carefully measured the orientation of the wafer by X-ray diffraction
and found the surface normal to deviate from the $[001]$ direction by less
than $0.05^{\circ }$. Such a small deviation from $[001]$ in a cubic
material is unlikely to be responsible for the observed anisotropies.
Density gradients, on the other hand, are often present in modulation-doped
samples, such as ours. It is a by-product of the MBE growth procedure with
off-axis doping sources. This applies particularly to substrates that are
not rotated, as is often the case for ultra-high mobility specimens. From
sequential low $B$-field transport measurements around the perimeter of our
sample we find, indeed, a small density gradient with a density variation of 
$<1\%$ across the sample. The direction of the gradient is roughly along the
current direction of the contact configuration that shows maxima in $R_{xx}$%
. Such a gradient can in principle account for an anisotropy. 

At any given magnetic field, the high-density end of the sample is
always at a slightly higher filling factor than the low-density
end. In particular, the high-density part may reside at $\nu =5$,
whereas the low-density end may reside at $\nu =4$, both being
separated by a narrow stripe, in which the Fermi energy resides within
a few delocalized states between these two filling factors. When the
current direction is along the gradient, the two voltage probes along
the edge of the specimen may reside at two different filling factors:
one at $\nu =5$ the other at $\nu =4$. Instead of the usual
zero-resistance of the IQHE minimum one would measure a resistance of
$\Delta R_{xy}=(1/4-1/5)h/e^2\approx 1.3k\Omega$, which is close to
the resistance of $R_{xx}\approx 1.5k\Omega$, observed in the $9/2$
peak. As the $B$-field is slightly raised or lowered, the separating
stripe of delocalized states moves quickly towards the high-density or
low-density end of the sample, subjecting both voltage probes to the
same IQHE filling factor, and therefore showing the usual
zero-resistance state in $R_{xx}$.  As the $B$-field is swept and
subsequent filling factors move through, one would observe a sequence
of spikes and zero-resistance states, in which the spikes have a
resistance of $R_{xx}=(1/\nu _{low}-1/\nu _{high})h/e^2$, not unlike
what is observed in the lowest temperature trace of Fig.1. In the
other current direction, normal to the gradient, the separating stripe
cannot fall between voltage probes, the spikes having the above
resistance values are absent and the normal IQHE/FQHE pattern
emerges. This is an attractive scenario to explain much of our
observations. However, it has several severe short comings.

First, with a density variation of only $1\%$
across the specimen, a $1\%$ field change is sufficient to move the
separating stripe across the sample. Therefore, the width of the spikes
should be $<1\%$, but are found to be $\sim 5\%$. Second, on reversing the $B
$-field direction, according to our simple model, the peaks should reverse
sign, whereas the experiment (not shown) shows, that the sign of the spikes
is maintained. Finally, the spikes and associated anisotropies are only
observed for $N>2$ and are not observed for $N\leq 2$, whereas our model
would not discriminate between different Landau levels. For these reasons we
have severe doubts that a density gradient can be origin of the observed
spikes and their extreme anisotropies. 

As to intrinsic origins for the
phenomena, a novel many-particle states may be responsible for the
observations. Koulakov {\em et al}\thinspace\cite{ref11} and Moessner and Chalker\cite
{ref12} have proposed new ground states for a clean electron system in weak
magnetic fields in which $N>>1$ Landau levels are totally occupied and the
next Landau level is only partially filled. Their work suggests that new
electronic states, such as charge-density-waves (CDW) or domain structures
may be created, which could be the origin of the stark anisotropies. In
particular, Koulakov {\em et al}\thinspace\cite{ref11} propose the existence of a
``stripe phase'' around half-filling of higher Landau levels, whereas a
``bubble phase'' of electron puddles on a triangular grid would exist in the
flanks of the Landau levels. In the stripe phase the occupation of the
highest level alternates spatially between totally full and totally empty on
the length scale of a classical cyclotron orbit $r_c\sim (\hbar N/eB)^{1/2}$.
Electrical current runs along the edge of the stripes. Such a phase could be
responsible for the anisotropy of the transport properties that we observe.
However, to be detectable in our experiment, the spatial symmetry of the
sample needs to be broken and the stripes need to be preferentially aligned
in one direction. The slight density gradient across the specimen may be
sufficient to break the symmetry and pin the stripes, leading to a
macroscopic anisotropy of the transport properties around half-filling. The
triangular phase, on the other hand, may be related to the observation of
the features in the vicinity of quarter filling. A recently modeled liquid
crystal-like phase, proposed by Fradkin and Kivelson\cite{ref13} may also be
related to our observations. However, we are far from being able to make a
positive experimental identification of any such novel electronic state.

In conclusion, we have observed stark anisotropies in the magnetotransport
of two-dimensional electrons in higher Landau levels. Several of the
experimentally observed features resemble FQHE features known from transport
in lower Landau levels, although, at closer inspection, they appear
inconsistent with an interpretation in such term.

Acknowledgments: We would like to thank F. Tsui for performing the X-ray
diffraction measurements. All transport experiments were performed at the
Francis Bitter Magnet Lab in Cambridge, Massachusetts. RRD and DCT are
supported in part by AFSOR, DOE, and by the NSF.

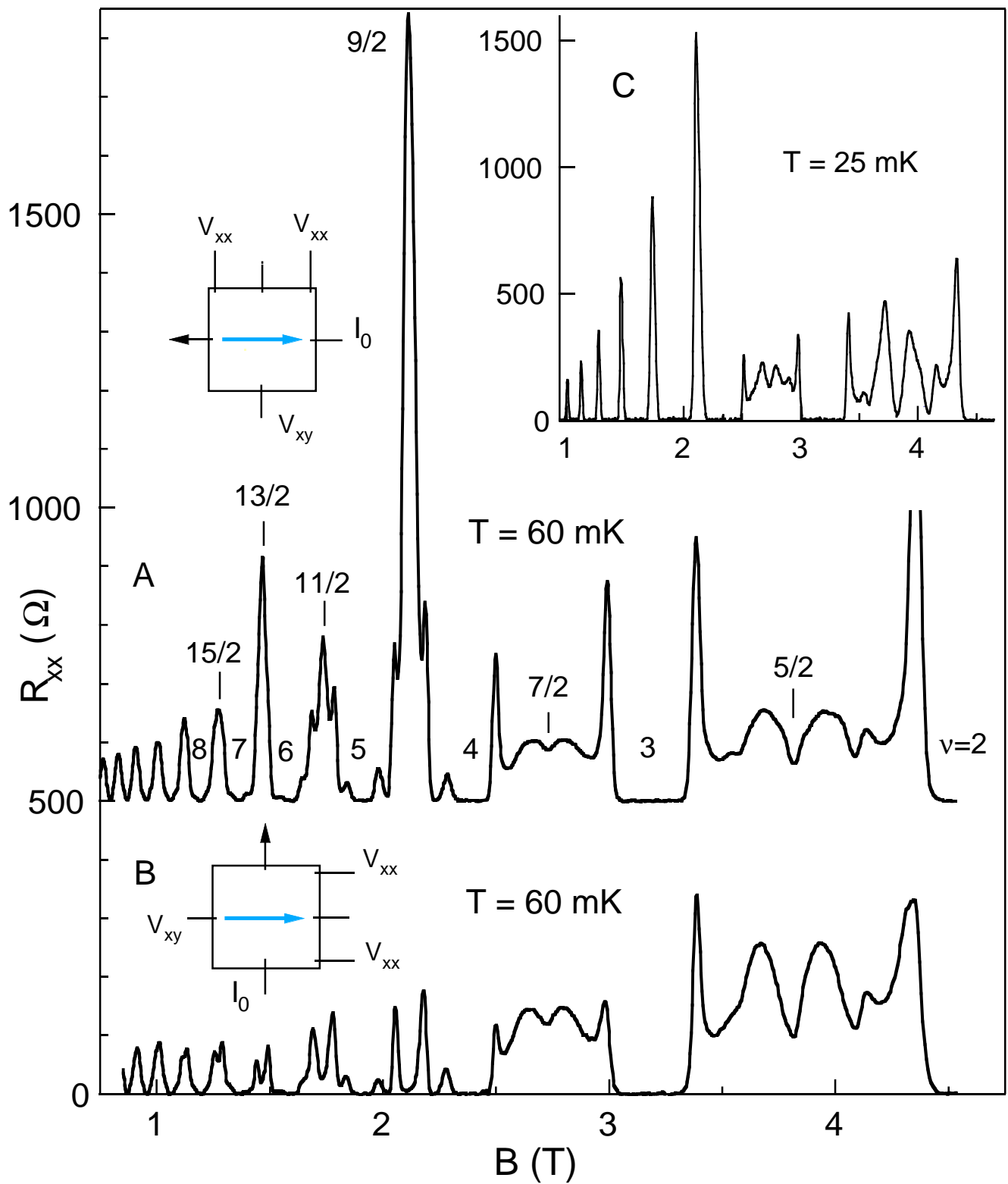
\begin{figure}[tbp]
\caption{ Magnetoresistance along two perpendicular in-plane directions of a
two-dimensional electron system at $60mK$. The schematics next to the traces 
$A$ and $B$ indicate the position of the voltage probes in the
magnetoresistance, $V_{xx}$, and Hall resistance , $V_{xy}$ (Fig.2)
configuration. $I_0$ identifies the current direction and the gray arrow on the
sample indicates the direction of a slight density gradient (see main text).
Strong anisotropies in the magnetoresistance are observed around
half-filling in higher Landau levels ($\nu =9/2$ through $\nu =17/2$). Trace 
$C$ of the insert shows $25mK$ data for the same contact configuration as
trace $A$.}
\label{Fig1}
\end{figure}

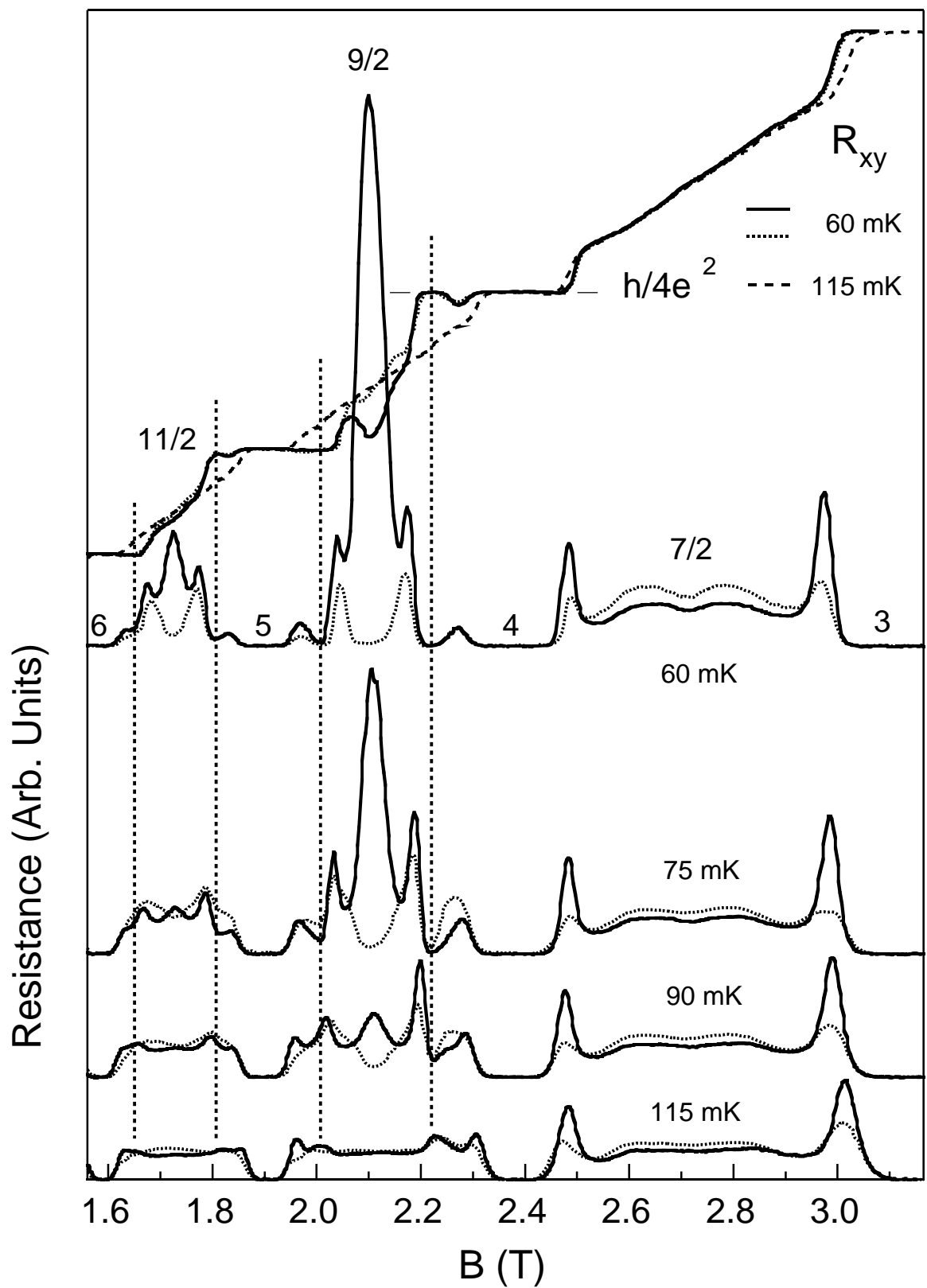
\begin{figure}[tbp]
\caption{Temperature dependence of a segment of data of Fig.1 The full lines
are from configuration $A$ , the dotted line from configuration $B$ of
Fig.1. The top traces are Hall resistances, $R_{xy}$, and their temperature
dependence. Around quarter-filling (dashed vertical lines), the
magnetoresistance develops minima not unlike FQHE minima, but the
concomitant $R_{xy}$ values approach IQHE plateaus.}
\label{Fig2}
\end{figure}

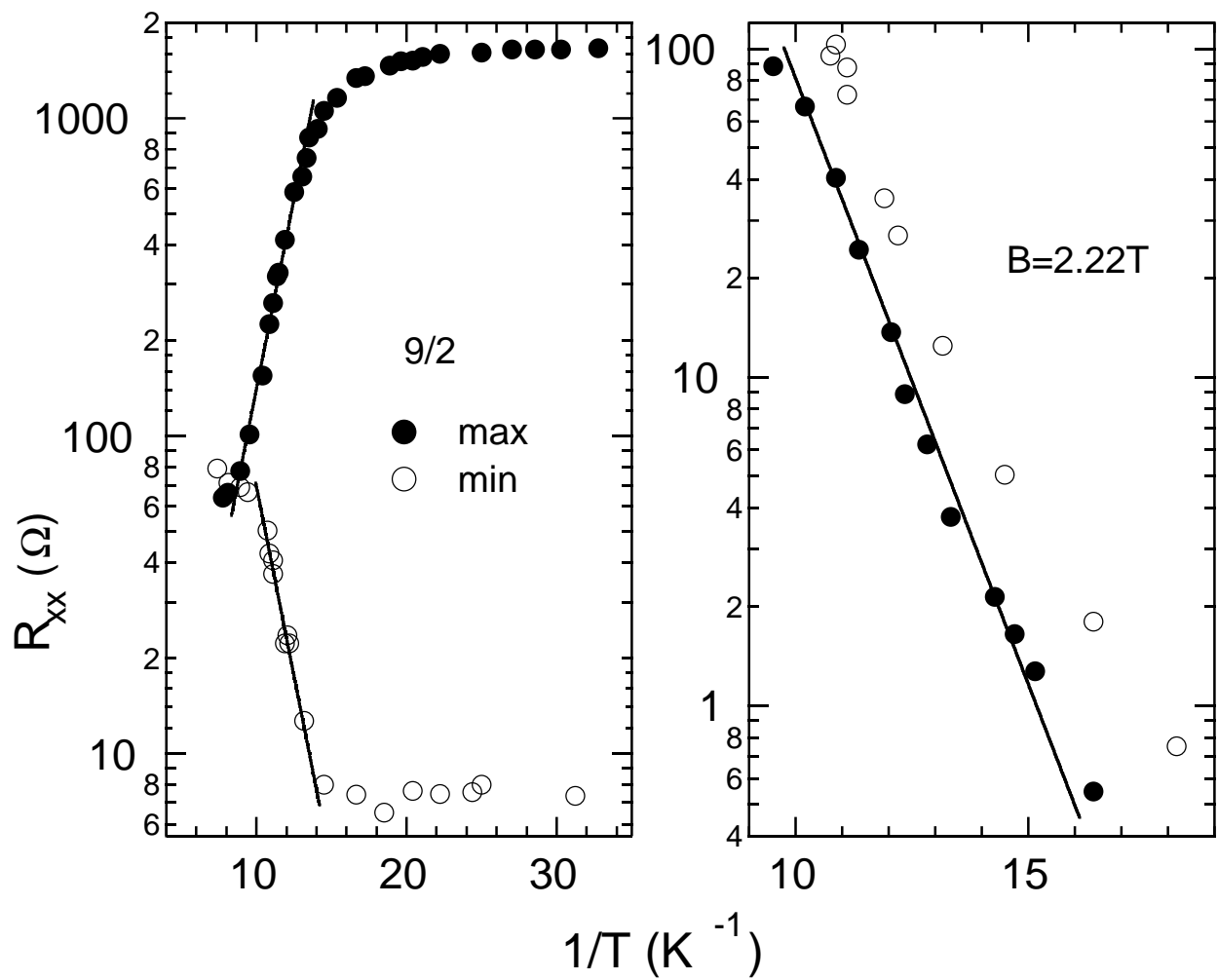
\begin{figure}[tbp]
\caption{ Left side: Temperature dependence of the $9/2$ resistance maximum
(filled circles) and $9/2$ resistance minimum (open circles) at $\nu =9/2$
versus inverse temperature. Although the behavior does not appear to be
thermally activated we indicate some characteristic energy scale of $E%
\sim0.55K$ by two straight line segments. Right side: Temperature
dependence of the magnetoresistance around $\nu\sim4+1/4$ $%
(B\sim2.22T)$ for both contact configurations of Fig.1. The behavior is
isotropic and thermally activated over two orders of magnitude with an
activation energy of $\Delta\sim1.7K$ (solid line).}
\label{Fig3}
\end{figure}


\begin{references}
\bibitem{ref1}  {\em Perspectives in Quantum Hall Effect-Novel Quantum
Liquids in Low-Dimensional Semiconductor Structures}, ed. by S. Das Sarma
and A. Pinczuk, Wiley and Sons, New York, 1997.

\bibitem{ref2}  {\em The Quantum Hall Effects}, T. Chakraborty and P.
Pietilainen, 85 Springer Series in Solid State Sciences, 1995

\bibitem{ref3}  R. L. Willett, J. P. Eisenstein, H. L. Stormer, D. C. Tsui,
A. C. Gossard, and J. H. English, Phys. Rev. Lett. {\bf 59}, 1776 (1987)

\bibitem{ref4}  R. G. Clark, R. J. Nicholas, J. R. Mallett, A. M. Suckling,
A. Usher, J. J. Harris, and C. T. Foxon, Proc. ICPS VIII, Stockholm, 1986,
O. Engstrom, edt., World Scientific, Singapor, 1987, p 393

\bibitem{ref5}  J. P. Eisenstein, R. L. Willett, H. L. Stormer, D. C. Tsui,
A. C. Gossard, and J. H. English, Phys. Rev. Lett. {\bf 66}, 997 (1988)

\bibitem{ref6}  P. L. Gammel, D. J. Bishop, J. P. Eisenstein, J. H. English,
A. C. Gossard, R. Ruel, and H. L. Stormer, Phys. Rev. {\bf B38}, 10128 (1988)

\bibitem{ref7}  H. L. Stormer, R. R. Du, D. C. Tsui, L. N. Pfeiffer, and K.
W. West, Bull. Amer. Phys. Soc. {\bf 38}, 235 (1993)

\bibitem{ref8}  M. P. Lilly, K. B. Cooper, J. P. Eisenstein, L. N. Pfeiffer,
and K. W. West, cond-mat/9808227

\bibitem{ref9}  G. Ebert, K. von Klitzing, C. Probst, E. Schubert, G.
Weiman, and W. Schlapp, J. Phys. C {\bf 17}, L775 (1984)

\bibitem{ref10}  R. G. Clark, R. J. Nicholas, A. Usher, C. T. Foxon and J.
J. Harris, Surf. Science {\bf 170}, 141 (1986)

\bibitem{ref11}  A. A. Koulakov, M. M. Fogler, and B. I. Shklovskii, Phys.
Rev. Lett. {\bf 76}, 499 (1996)

\bibitem{ref12}  R. Moessner and J. T. Chalker, Phys. Rev. {\bf B54}, 5006
(1996)

\bibitem{ref13}  E. Fradkin and S. A. Kivelson, cond-mat/9810148
\end{references}
\end{document}